\documentclass[aps,prl,twocolumn,showpacs,groupedaddress]{revtex4}

\usepackage{graphicx}
\usepackage{dcolumn}
\usepackage{bm}
\usepackage{amssymb}

\usepackage[usenames]{color}

\begin{document}

\newcommand{\dzero}     {D0}
\newcommand{\met}       {\mbox{$\not\!\!E_T$}}
\newcommand{\deta}      {\mbox{$\eta^{\rm det}$}}
\newcommand{\meta}      {\mbox{$\left|\eta\right|$}}
\newcommand{\mdeta}     {\mbox{$\left|\eta^{\rm det}\right|$}}
\newcommand{\rar}       {\rightarrow}
\newcommand{\rargap}    {\mbox{ $\rightarrow$ }}
\newcommand{\tbbar}     {\mbox{$tb$}}
\newcommand{\tqbbar}    {\mbox{$tqb$}}
\newcommand{\ttbar}     {\mbox{$t\bar{t}$}}
\newcommand{\bbbar}     {\mbox{$b\bar{b}$}}
\newcommand{\ccbar}     {\mbox{$c\bar{c}$}}
\newcommand{\qqbar}     {\mbox{$q\bar{q}$}}
\newcommand{\ppbar}     {\mbox{$p\bar{p}$}}
\newcommand{\comphep}   {\sc{c}\rm{omp}\sc{hep}}
\newcommand{\herwig}    {\sc{herwig}}
\newcommand{\pythia}    {\sc{pythia}}
\newcommand{\alpgen}    {\sc{alpgen}}
\newcommand{\singletop} {\rm{SingleTop}}
\newcommand{\reco}      {\sc{reco}}
\newcommand{\Mchiggs}   {\mbox{$M({\rm jet1,jet2},W)$}}

\lefthyphenmin=6
\righthyphenmin=6

\title{Search for anomalous $\boldmath{Wtb}$ couplings in single top quark production} 
%
\author{V.M.~Abazov$^{36}$}
\author{B.~Abbott$^{75}$}
\author{M.~Abolins$^{65}$}
\author{B.S.~Acharya$^{29}$}
\author{M.~Adams$^{51}$}
\author{T.~Adams$^{49}$}
\author{E.~Aguilo$^{6}$}
\author{M.~Ahsan$^{59}$}
\author{G.D.~Alexeev$^{36}$}
\author{G.~Alkhazov$^{40}$}
% ...

\affiliation{\vspace{0.1 in}(The D\O\ Collaboration)\vspace{0.1 in}}
% ...
\affiliation{$^{6}$University of Alberta, Edmonton, Alberta, Canada,
                Simon Fraser University, Burnaby, British Columbia, Canada,
                York University, Toronto, Ontario, Canada, and
                McGill University, Montreal, Quebec, Canada}
% ...
\affiliation{$^{29}$Tata Institute of Fundamental Research, Mumbai, India}
% ...
\affiliation{$^{36}$Joint Institute for Nuclear Research, Dubna, Russia}
% ...
\affiliation{$^{40}$Petersburg Nuclear Physics Institute,
                St. Petersburg, Russia}
% ...
\affiliation{$^{49}$Florida State University, Tallahassee, Florida 32306, USA}
% ...
\affiliation{$^{51}$University of Illinois at Chicago,
                Chicago, Illinois 60607, USA}
% ...
\affiliation{$^{59}$Kansas State University, Manhattan, Kansas 66506, USA}
% ...
\affiliation{$^{65}$Michigan State University,
                East Lansing, Michigan 48824, USA}
% ...
\affiliation{$^{75}$University of Oklahoma, Norman, Oklahoma 73019, USA}
% ...

\begin{abstract}
 In 0.9~fb$^{-1}$ of $p \bar p$ collisions, D0 has observed an excess of events with an isolated lepton, missing transverse momentum, and two to four jets. This excess is consistent with single top quark production. We examine these data to study the Lorentz structure of the $Wtb$ coupling. The standard model predicts a left-handed vector coupling at the $Wtb$ vertex. The most general lowest dimension, $CP$-conserving Lagrangian admits right-handed vector and left- or right-handed tensor couplings as well. We find that the data prefer the left-handed vector coupling and set upper limits on the anomalous couplings. These are the first direct constraints on a general $Wtb$ interaction and the first direct limits on left- and right-handed tensor couplings.

\pacs{14.65.Ha; 12.15.Ji; 13.85.Qk}

\end{abstract}
\maketitle

\vspace{-0.1in}
Recently, we presented evidence for single top quark production in $p \bar p$ collisions at $\sqrt s=1.96$~TeV~\cite{abazov:181802} based on 0.9~fb$^{-1}$ of data collected using the {\dzero} detector~\cite{UpgradedD0:2006} at the Fermilab Tevatron collider. In this Letter, we report an extension of this analysis using the same data set and similar analysis tools to study the consistency of this excess with different hypotheses for the couplings involved in single top quark production. This is the first time such a test has been carried out.

The standard model (SM) has been extraordinarily successful in describing the data taken at the energies of present colliders. However, we know that the electroweak symmetry breaking sector of the SM gives rise to many unanswered questions, making a strong case for new physics beyond the SM. This new physics can manifest itself in the production of new particles or in corrections to SM processes that change the effective couplings of SM particles. The interactions between quarks and  gauge bosons have been measured precisely at the CERN Large Electron Positron collider~\cite{LEP_status} except for the top quark, which was not kinematically accessible. The large mass of the top quark has prompted speculation that the top quark may play a special role in the mechanism of electroweak symmetry breaking and thus have non-standard interactions with weak gauge bosons.
 We can probe the interactions of top quarks with $W$ bosons via measurements of single top quark production and top quark decays in $t\bar{t}$ production, each yielding complementary information.

\begin{figure}[!h!btp]
\includegraphics[width=0.35\textwidth]
{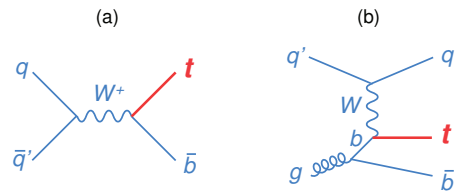}
\vspace{-0.1in}
\caption[feynman]{Feynman diagrams for (a) $s$-channel
and (b) $t$-channel single top quark production.}
\label{feynman-diagrams}
\end{figure}

The dominant tree level Feynman diagrams for single top quark production in $p \bar p$ collisions are illustrated in Fig.~\ref{feynman-diagrams}. We use the notation ``$tb$" for the sum of the $s$-channel processes $t\bar{b}$ and $\bar{t}b$ and ``$tqb$" for the sum of the $t$-channel processes $tq\bar{b}$ and $\bar{t}\bar{q}b$. We assume that single top quark production proceeds exclusively through $W$ boson exchange. Therefore, extensions of the SM in which single top quarks are produced via flavor-changing neutral current interactions~\cite{run2-d0-fcnc} or the exchange of new massive scalar~\cite{run2-d0-chargedhiggs} or vector bosons~\cite{run2-d0-wprime}, are not considered here.  
We further assume that $|V_{td}|^2+|V_{ts}|^2 \ll |V_{tb}|^2$, i.e., the $Wtb$ vertex dominates top quark production and decay.
 Finally, we assume that  the $Wtb$ vertex is $CP$ conserving.

The most general, lowest dimension, $CP$ conserving, Lagrangian for the $Wtb$ vertex is~\cite{cpyuan_0503040v3}:
\begin{eqnarray*}
\mathcal{L} &=&\frac{g}{\sqrt{2}}W^-_{\mu}{\bar{b}}\gamma^\mu \left(f^{L}_1 P_L + f^{R}_1 P_R\right)t \\
&-&\frac{g}{\sqrt{2} M_W} {\partial _{\nu}} W^-_{\mu}{\bar{b}} \sigma^{\mu\nu} \left(f^{L}_2 P_L + f^{R}_2 P_R\right)t   + h.c., 
\label{coupling}
\end{eqnarray*}
 where  $M_W$ is the mass of $W$ boson, $P_{L}=(1 - \gamma_5)/2$ is the left-handed projection operator and $P_R=(1 + \gamma_5)/2$ is the right-handed projection operator. In the SM the values of the form factors are  $f^{L}_{1} \approx 1 $, $f^{L}_{2}=f^{R}_{1}=f^{R}_{2} =0$.
In this case the predicted cross section for single top quark production is $2.9\pm0.3$~pb~\cite{singletop-xsec-sullivan}. 

\begin{figure}[!h!tbp]
\includegraphics[width=0.43\textwidth]{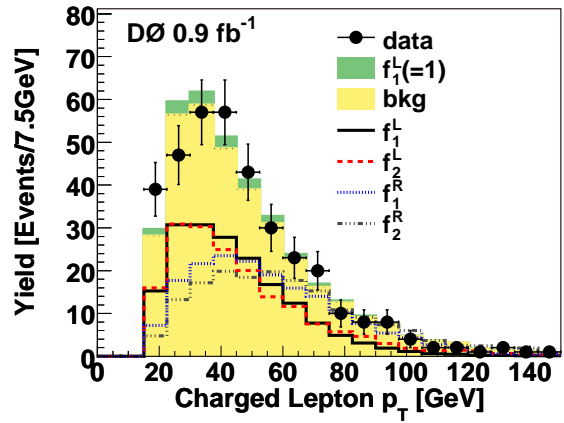}
\vspace{-0.1in}
\caption[lep-pt]{Charged lepton $p_T$ spectrum from data and
expectation for SM single top production plus background for
events with two jets, one $b$-tagged jet. Superimposed are the
 distributions from single top quark production with different couplings (all other couplings set to zero) normalized to ten times the SM single top quark cross section.}
\label{lep-pt_sum}
\vspace{-.1in}
\end{figure}

The presence of anomalous couplings can change angular distributions and event kinematics as demonstrated by the $p_T$ spectrum of the charged lepton from the decay of the top quark in Fig.~\ref{lep-pt_sum}. Such differences can be used to distinguish these couplings~\cite{dudko-boos,Wtb-coupling-yuan1}.
The magnitude of the right-handed vector coupling and tensor couplings can be indirectly constrained by the measurement of the $b \rightarrow s\gamma$ branching fraction~\cite{bsgamma}. Direct constraints on the combination of several couplings can be obtained  from the measurement of the $W$ boson helicity in top quark decays~\cite{Whel-d0}.
The predicted single top quark production cross sections are $2.7\pm0.3$~pb if $f^{R}_{1}=1$ and $10.4\pm1.4$~pb if $f^{L}_{2}=1$ or $f^{R}_{2}=1$ and the other couplings vanish~\cite{dudko-boos}.

Ideally,  we would like to set limits on all four couplings, $f^{L}_{1}$, $f^{L}_{2}$, $f^{R}_{1}$ and $f^{R}_{2}$,  simultaneously. This, however, requires more data than is currently available. We therefore look at two couplings at a time and assume that the other two are negligible. We consider three cases in which we allow the left-handed vector coupling $f^{L}_{1}$ and any one of the three non-standard couplings to be non-zero. We refer to these as $(L_1,L_2)$, $(L_1,R_1)$, and $(L_1,R_2)$.


We look for events in which the top quark decays to a $W$ boson and a $b$ quark, followed by the decay of the $W$ boson to an electron or a muon, and a neutrino. The event selection is the same as in Ref.~\cite{abazov:181802}. 
To enhance the signal content of the selected data sample, one or two of the jets are required to be identified as originating from long-lived $b$~hadrons~\cite{Scanlon}.


We model the single top quark signal using the {\sc comphep-singletop}  Monte Carlo event generator~\cite{singletop-mcgen} and the anomalous $Wtb$ couplings are considered in both production and decay in the generated signal samples. The event kinematics for both $s$-channel and $t$-channel reproduce distributions from next-to-leading-order calculations~\cite{singletop-xsec-sullivan}. The decay of the top quark and the resulting $W$~boson are carried out in the {\sc singletop} generator in order to preserve the information about the spin of the particles. {\sc pythia}~\cite{pythia} is used to add the underlying event, initial and final-state radiation, and for hadronization. The top quark mass is set to 175~GeV and the CTEQ6L1 parton distribution functions~\cite{cteq} are used.

Background contributions from $W+$jets and {\ttbar} production are
simulated  using the {\alpgen} leading-order Monte Carlo event generator~\cite{alpgen} 
interfaced to {\sc pythia}. A parton-jet matching algorithm~\cite{jet-matching} is used to avoid double counting. The response of the D0 detector to the Monte Carlo events is simulated using {\sc geant}~\cite{geant}. Simulated events are processed through the same reconstruction software used for data and efficiencies and resolutions are corrected to match the performance of the reconstruction for data. The {\ttbar} background is normalized using the theoretical cross section~\cite{ttbar-xsec-1}. The multijet background is modeled using events from data containing nonisolated leptons that otherwise resemble the signal events. The $W$+jets background is normalized such that the number of events predicted by the simulation agrees with the number of events observed in each analysis channel (defined by lepton flavor and jet multiplicity) before $b$~tagging is applied.

After all cuts we select 1,398 $b$~tagged lepton+jets events, which we expect to contain $62\pm 13$ single top quark events, $348\pm80$ {\ttbar} events, $849\pm222$ $W+$jets events, and $202\pm48$ multijet events. Within each channel the signal efficiency of the complete selection does not depend strongly on the assumed $Wtb$ coupling. The selection efficiencies for signal with different $Wtb$ couplings vary between ($1.07\pm0.15$)\% and ($1.52\pm0.16$)\% for $tb$ events with 1 $b$~tag, between ($0.86\pm0.13$)\% and ($1.14\pm0.14$)\% for $tqb$ events with 1 $b$~tag, between ($0.40\pm0.08$)\% and ($0.60\pm0.10$)\% for $tb$ events with 2 $b$~tags, and between ($0.07\pm0.01$)\% and ($0.10\pm0.02$)\% for $tqb$ events with 2 $b$~tags.

Systematic uncertainties in the signal and background models are 
estimated using the methods described in Ref.~\cite{abazov:181802}. 
 The dominant contributions to the uncertainties in the background estimate
come from: the normalization of the {\ttbar} background (18\%), which
includes the top quark mass uncertainty; the
normalization of the $W$+jets and multijets backgrounds to data
(17\%--27\%), which includes the uncertainty in the fraction of events with heavy flavor production; and the $b$-tagging efficiencies (12\%--17\% for
double-tagged events). The
uncertainties from the jet energy scale corrections (1\%--20\%) and the
$b$~tagging probabilities affect both the shape and normalization of
the simulated distributions. All other components contribute at the few percent level.



We use boosted decision trees~\cite{decisiontrees-breiman,boosting-freund} to  discriminate between the single top quark signal and background.
For each of the three coupling scenarios, we train trees in four analysis channels defined by lepton flavor and $b$~tag multiplicity. For each scenario the signal samples consist of a sample of events generated with left-handed vector coupling set to one, i.e. with SM coupling,  and a sample of events generated with the non-standard coupling set to one and all other couplings set to zero. The background sample consists of events from all background sources in the expected proportions. 

\begin{figure*}[!h!tbp]
\includegraphics[width=0.3\textwidth]{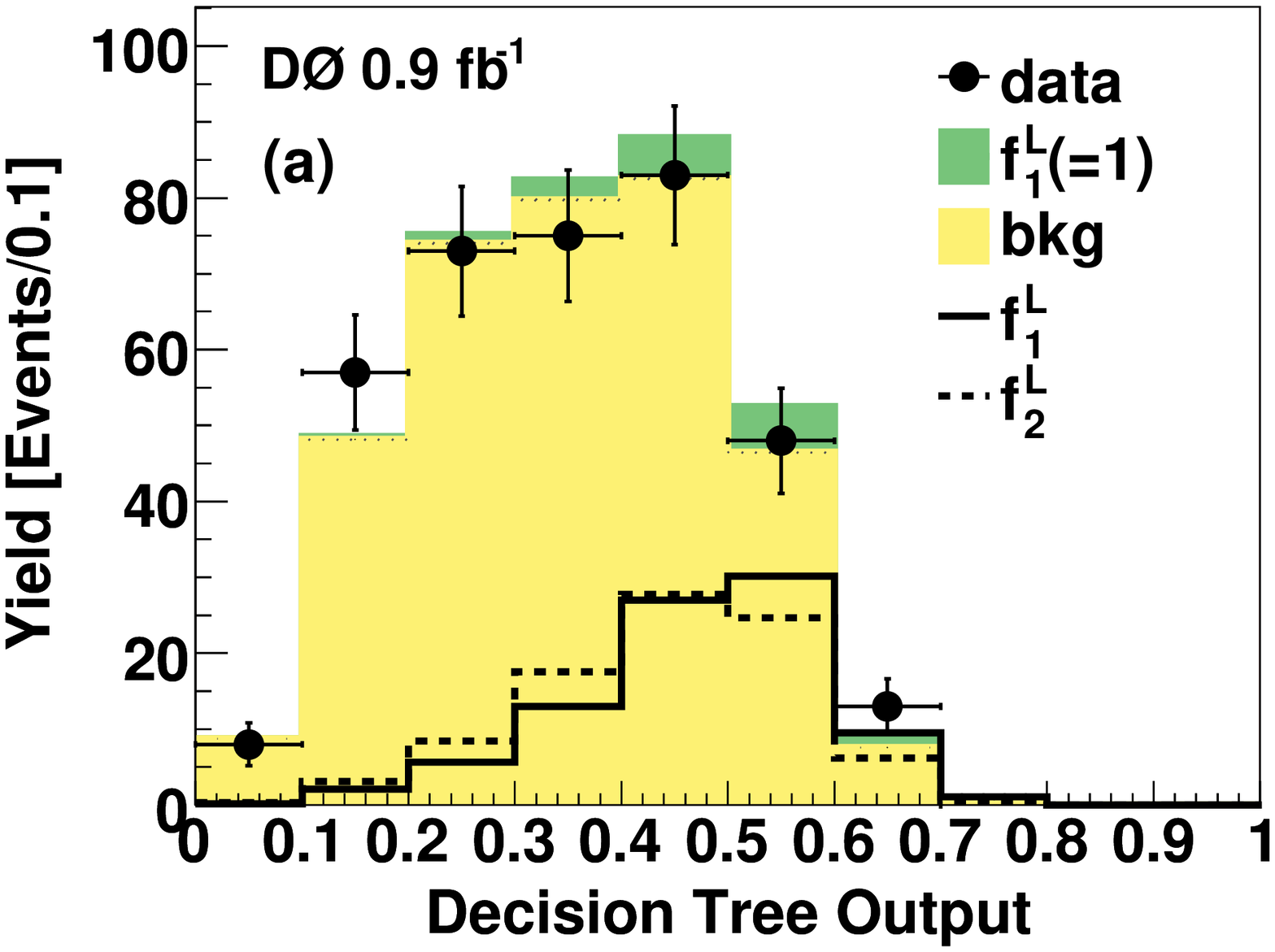}
\includegraphics[width=0.3\textwidth]{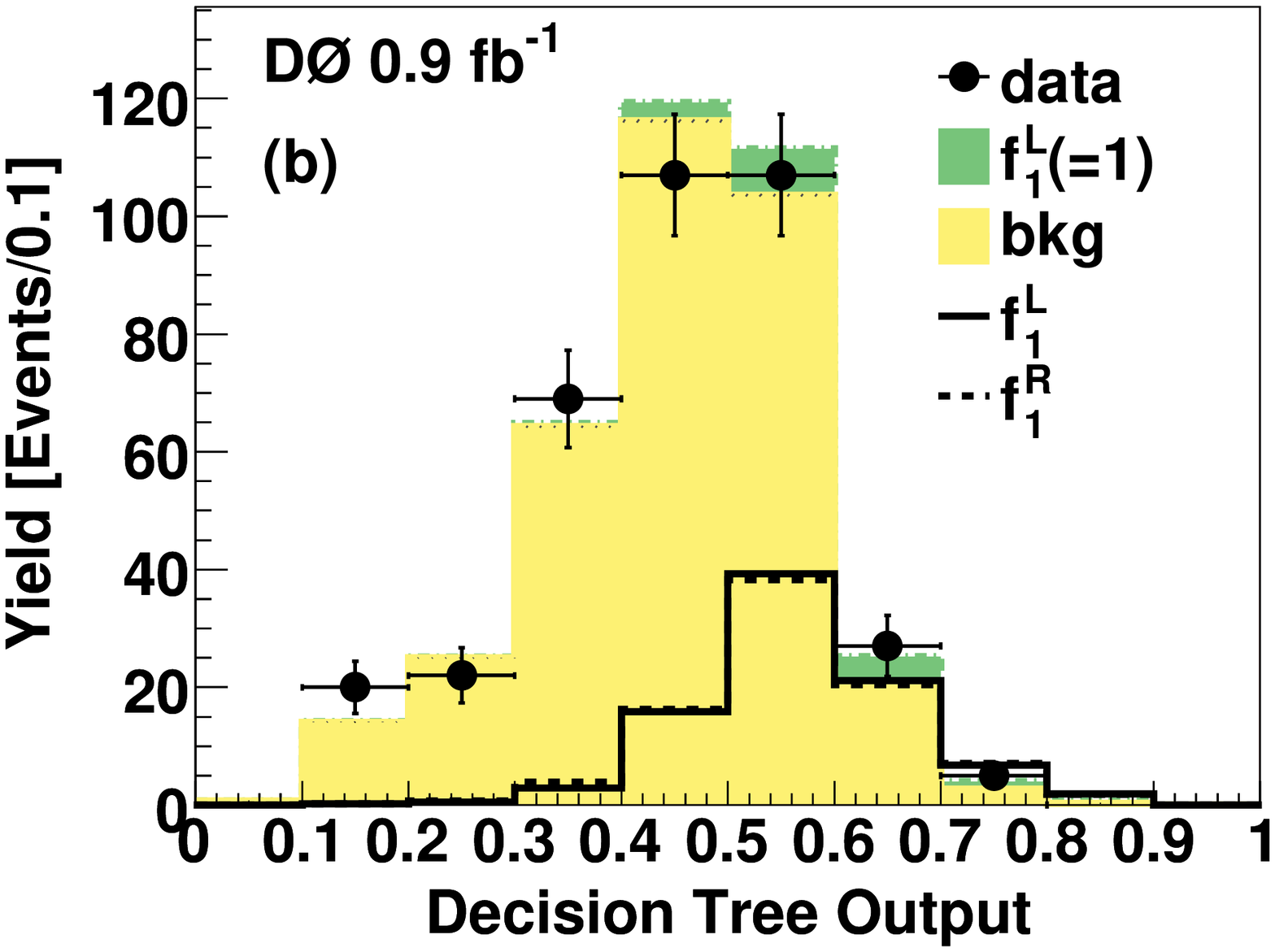}
\includegraphics[width=0.3\textwidth]{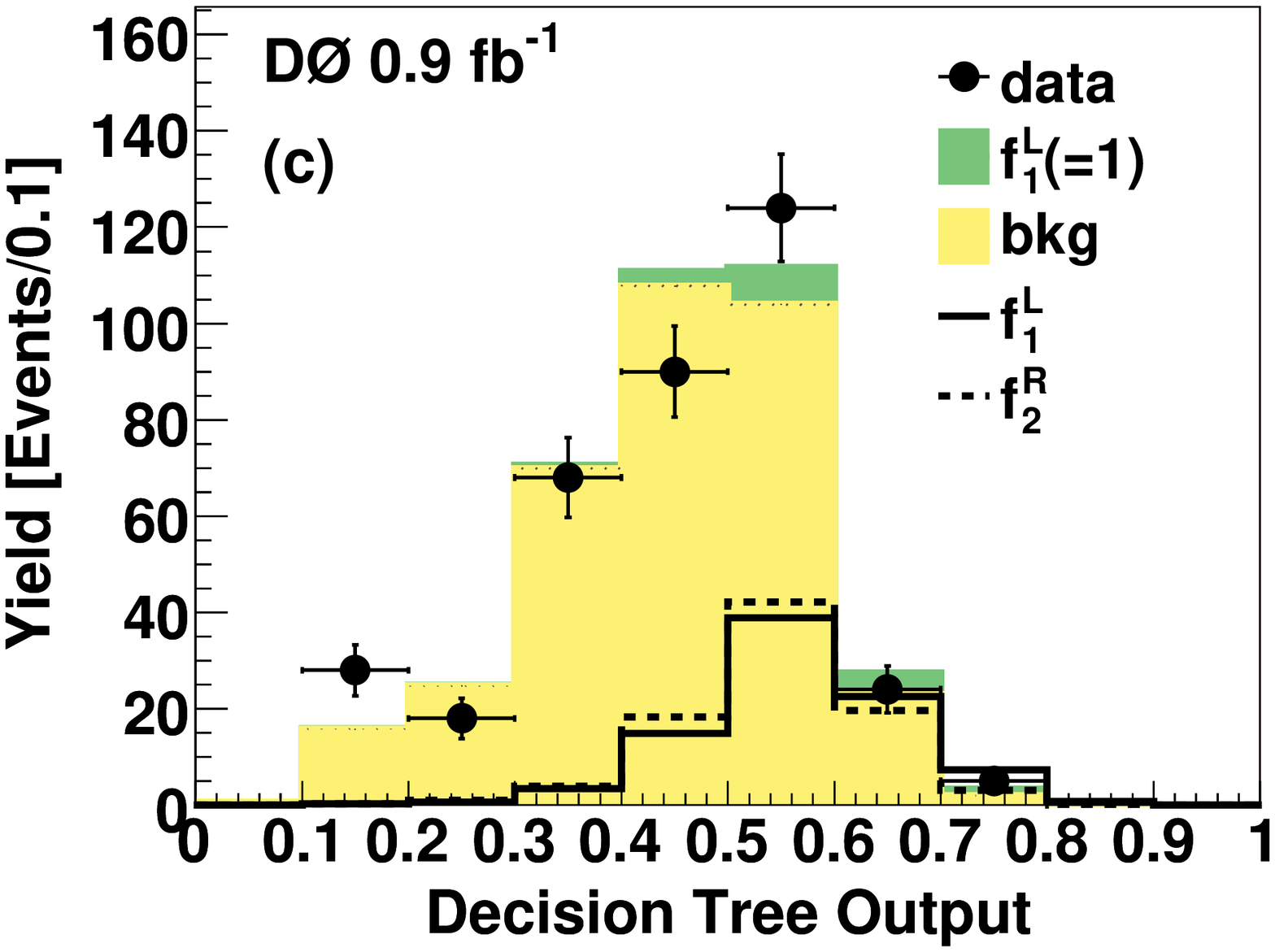}
\vspace{-0.1in}
\caption[dtoutputs]{Boosted decision tree output distributions for data and sum of SM signal and  backgrounds for events with two jets and one $b$-tagged jet for (a) the ($L_1$,$L_2$) scenario,  (b) the ($L_1$,$R_1$) scenario, and (c) the ($L_1$,$R_2$) scenario. Superimposed are the distributions for the single top quark signals with different couplings normalized to five times the SM single top quark cross section.}
\label{dtoutputs}
\vspace{-.1in}
\end{figure*}

We use 50 variables in the training, the 49 variables that were used in Ref.~\cite{abazov:181802} plus the lepton $p_T$ which helps distinguish the signals with different couplings, as can be seen in Fig.~\ref{lep-pt_sum}. The variables describe individual object kinematics, global event kinematics, and angular correlations. The boosted decision trees produce a continuous output distribution ranging from zero to one, with background tending closer to zero and signal tending closer to one. Figure~\ref{dtoutputs} shows  representative output distributions for the data and the sum of SM signal and backgrounds for the electron channel with two jets and one $b$-tagged jet in each of the three anomalous coupling scenarios.


\begin{figure}[!h!tbp]
\includegraphics[width=0.23\textwidth]{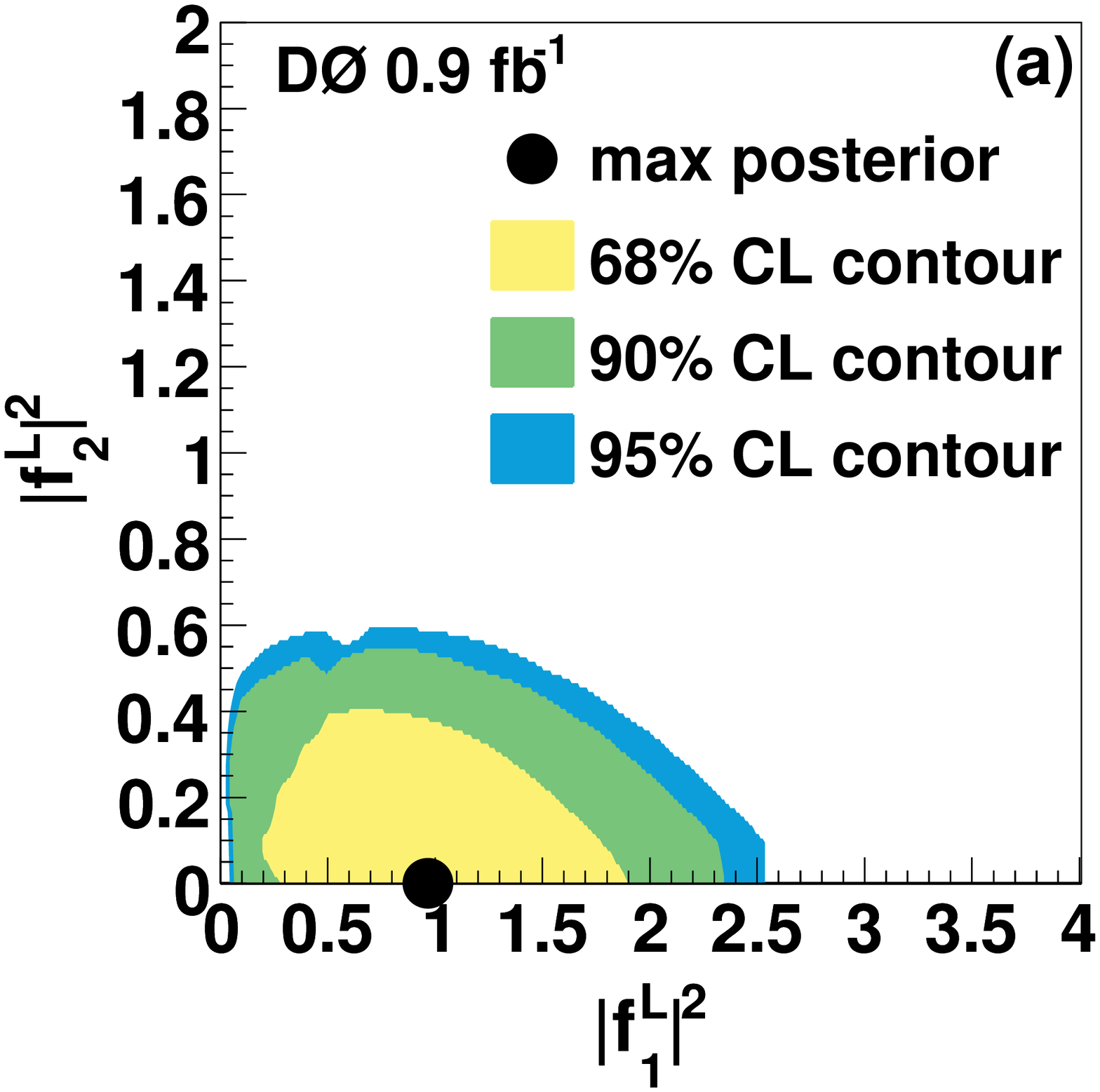}
\includegraphics[width=0.23\textwidth]{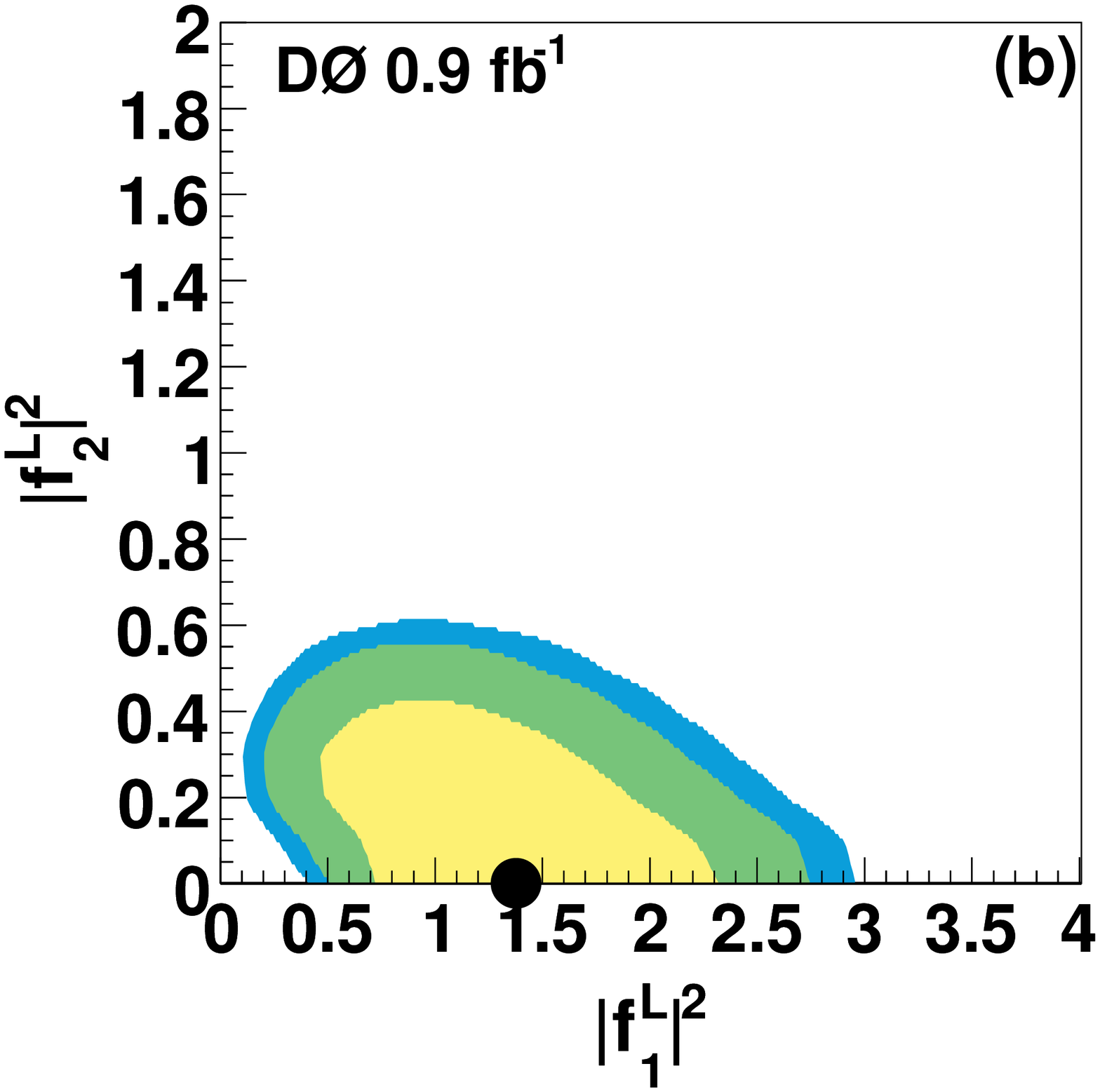}\\
\includegraphics[width=0.23\textwidth]{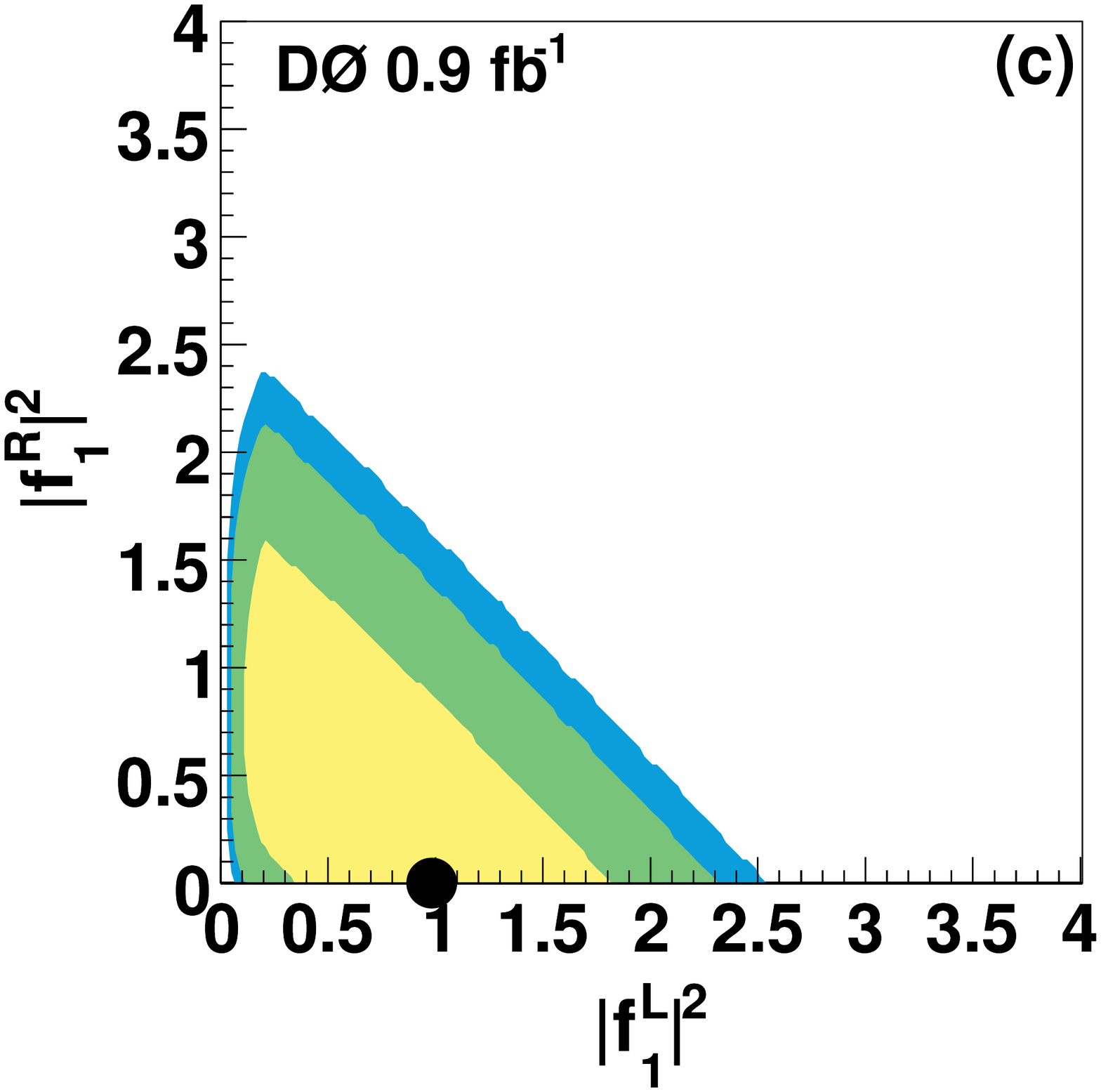}
\includegraphics[width=0.23\textwidth]{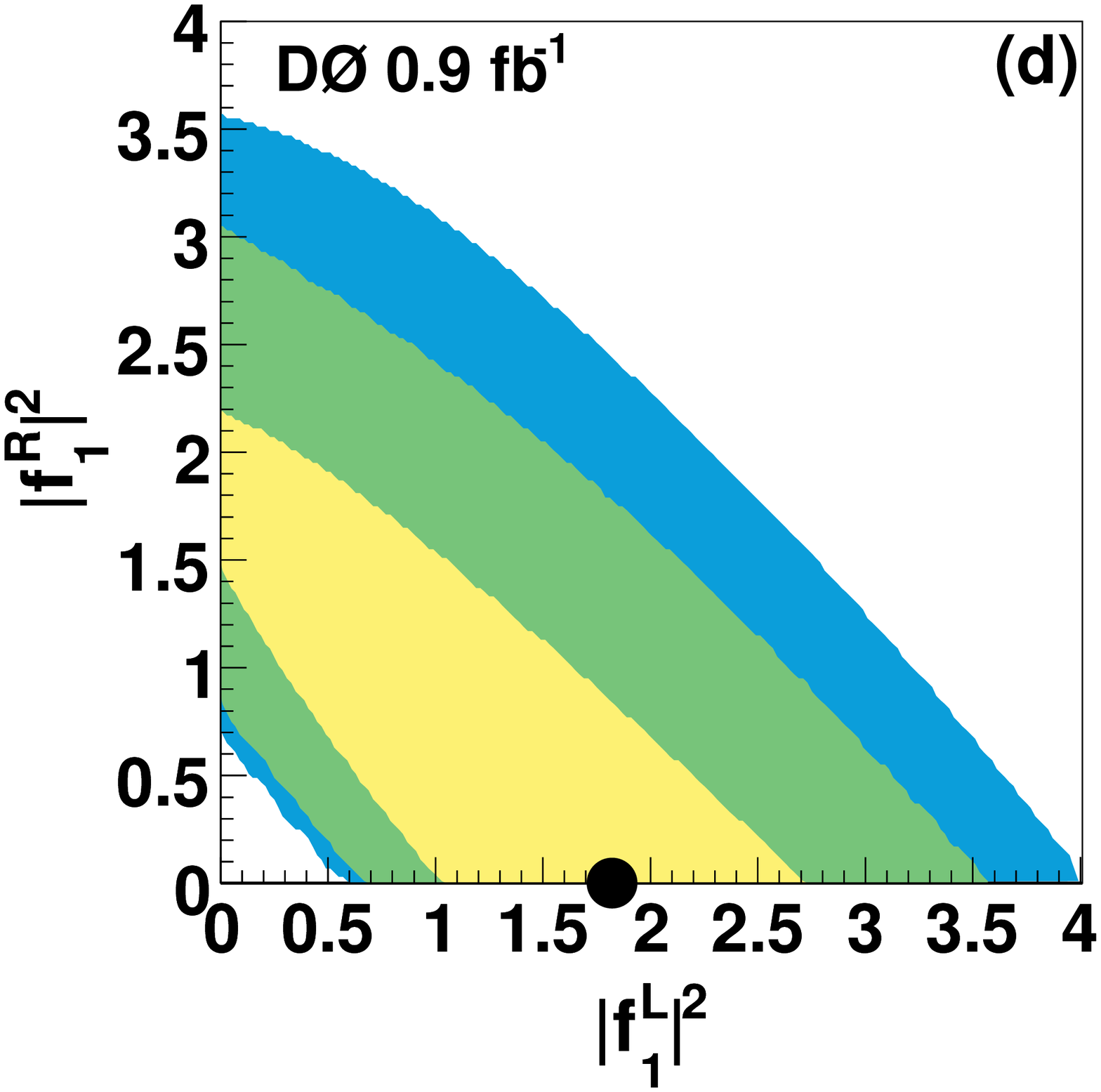}\\
\includegraphics[width=0.23\textwidth]{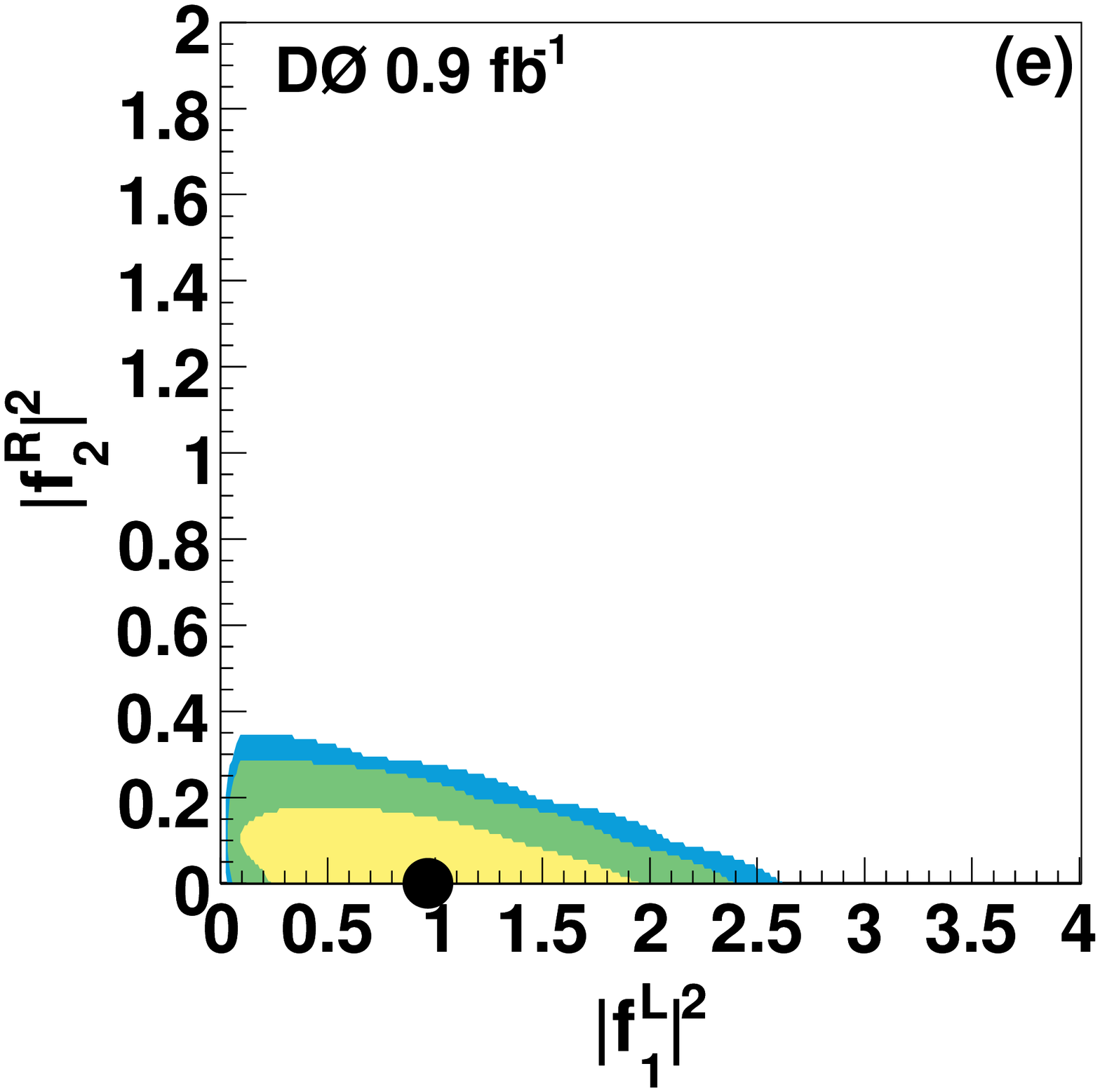}
\includegraphics[width=0.23\textwidth]{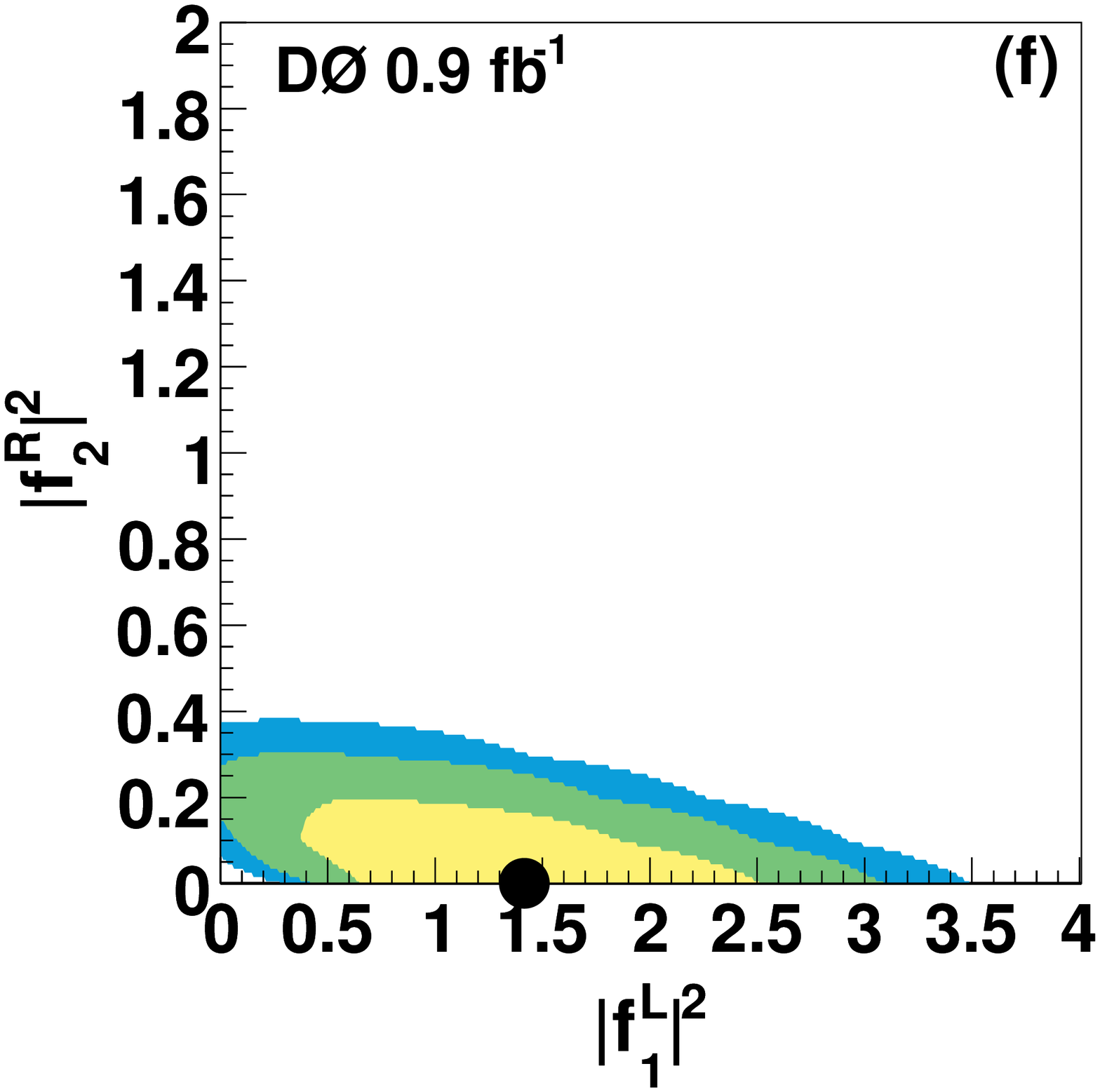}
\caption[measfullsys_2D_0100]{Plots of the two-dimensional posterior probability density for the anomalous couplings. The plots on the left show the expectation for left-handed vector couplings and the plots on the right show the observed posterior from our data. The upper row (a, b) shows the plots for the ($L_1$,$L_2$) scenario, the middle row (c, d) for the ($L_1$,$R_1$) scenario, and the bottom row  (e, f) for the ($L_1$,$R_2$) scenario.}
\label{fig:measfullsys_2D}
\vspace{-.1in}
\end{figure}

We use Bayesian statistics~\cite{bayes-limits} to compare the output distribution of the decision trees from data to expectations for single top quark production. For any pair of values of the two couplings that are considered non-zero, we compute the expected output distribution by superimposing the distributions from the two signal samples  with the non-standard coupling and from the background samples in the appropriate proportions. In case of the $(L_1,L_2)$ scenario, the two amplitudes interfere, and we use a superposition of three signal samples, one with left-handed vector couplings, one with the left-handed tensor coupling only set to one, and one with both couplings set to one to take into account the effect of the interference. We then compute a likelihood as a product over all bins and channels. Here we use twelve channels defined by lepton flavor, $b$~tag multiplicity, and jet multiplicity (2, 3, or 4). We assume Poisson distributions for the observed counts, and flat nonnegative prior probabilities for the signal cross sections. The prior for the combined signal acceptance and background yields is a multivariate Gaussian with uncertainties and correlations described by a covariance matrix. A two-dimensional posterior probability density is computed as a function of $|f^{L}_{1}|^2$ and $|f_X|^2$, where $f_X$ is any of the other three non-standard couplings, in each channel. These probability distributions are shown in~Fig.~\ref{fig:measfullsys_2D}. We quote the values of the couplings that maximize the two-dimensional likelihood as our measurements. In all three scenarios we measure zero for the right-handed vector, and left- and right-handed tensor couplings.
 We compute 95\% C.L. upper limits on these couplings by integrating out the left-handed vector coupling to get a one-dimensional posterior probability density. The measured values are given in Table~\ref{table:obslim}. The data favor the left-handed vector hypothesis over the alternative hypotheses.

In summary, we have studied the excess observed in 0.9~fb$^{-1}$ of  {\dzero} data in the search for single top quark production. We attribute this excess to single top quark production and study its consistency with different hypotheses for the structure of the $Wtb$ coupling and find that the data prefer the left-handed vector coupling over the alternative hypotheses studied. These are the first direct constraints on a general $Wtb$ interaction and the first direct limits on left- and right-handed tensor couplings.


\begin{table}[t!!!]
\caption{\label{table:obslim} Measured values of the total cross section for single top production and one-dimensional limits on $Wtb$ couplings in the three scenarios.
}
\begin{ruledtabular}
\begin{tabular}{lcl}
Scenario  & Cross Section  &  \multicolumn{1}{c}{Coupling}   \\
\hline
$(L_1,L_2)$  & $4.4^{+2.3}_{-2.5}$~pb & $|f^{L}_{1}|^2=1.4 ^{+0.6}_{-0.5}$ \\
             &                        & $|f^{L}_{2}|^2<0.5$ at 95\% C.L.   \\
$(L_1,R_1)$  & $5.2^{+2.6}_{-3.5}$~pb & $|f^{L}_{1}|^2=1.8 ^{+1.0}_{-1.3}$ \\
             &                        & $|f^{R}_{1}|^2<2.5$ at 95\% C.L.  \\
$(L_1,R_2)$  & $4.5^{+2.2}_{-2.2}$~pb & $|f^{L}_{1}|^2=1.4 ^{+0.9}_{-0.8}$ \\
             &                        & $|f^{R}_{2}|^2<0.3$ at 95\% C.L.   \\
\end{tabular}
\end{ruledtabular}
\vspace{-.1in}
\end{table}

%
We thank the staffs at Fermilab and collaborating institutions, 
and acknowledge support from the 
DOE and NSF (USA);
CEA and CNRS/IN2P3 (France);
FASI, Rosatom and RFBR (Russia);
CNPq, FAPERJ, FAPESP and FUNDUNESP (Brazil);
DAE and DST (India);
Colciencias (Colombia);
CONACyT (Mexico);
KRF and KOSEF (Korea);
CONICET and UBACyT (Argentina);
FOM (The Netherlands);
STFC (United Kingdom);
MSMT and GACR (Czech Republic);
CRC Program, CFI, NSERC and WestGrid Project (Canada);
BMBF and DFG (Germany);
SFI (Ireland);
The Swedish Research Council (Sweden);
CAS and CNSF (China);
and the
Alexander von Humboldt Foundation (Germany).
%
\vspace{-0.1in}


\end{document}